%% file: QuantumLT_9.tex
\documentclass[aps,prl,twocolumn,superscriptaddress]{revtex4-1}

\usepackage{graphicx}  
\usepackage{amsmath} 

\usepackage{graphicx}
\usepackage{amsmath,amssymb,amsfonts}
\usepackage{textcomp}
\usepackage{hyperref}
\usepackage{gensymb}
\usepackage{verbatim}
\usepackage{braket}

\hypersetup{breaklinks=true,colorlinks=true,urlcolor=black}
\usepackage{color}
\usepackage{soul}
\usepackage{gensymb}

\DeclareGraphicsExtensions{.jpg,.pdf,.png,.eps}

\begin{document}


\title{Quantum Liouville's theorem based on Haar measure}



\author{B.Q.~Song}
\affiliation{Ames National Laboratory, Iowa State University, Ames, Iowa 50011, USA}
\affiliation{Department of Physics and Astronomy, Iowa State University, Ames, Iowa 50011, USA}
\author{J.D.H.~Smith}
\affiliation{Ames National Laboratory, Iowa State University, Ames, Iowa 50011, USA}
\affiliation{Department of Mathematics, Iowa State University, Ames, Iowa 50011, USA}
\author{L.~Luo}
\author{J.~Wang}
\affiliation{Ames National Laboratory, Iowa State University, Ames, Iowa 50011, USA}
\affiliation{Department of Physics and Astronomy, Iowa State University, Ames, Iowa 50011, USA}


\date{\today}

\begin{abstract}
Liouville theorem (LT) reveals robust incompressibility of distribution function in phase space, given arbitrary potentials. However, its quantum generalization, Wigner flow, is compressible, i.e., LT is only conditionally true (e.g., for perfect Harmonic potential). We develop quantum Liouville theorem (rigorous incompressibility) for arbitrary potentials (interacting or not) in Hamiltonians. Haar measure, instead of symplectic measure $dp{\wedge}dq$ used in Wigner’s scheme, plays a central role. The argument is based on general measure theory, independent of specific spaces or coordinates. Comparison of classical and quantum is made: for instance, we address why Haar measure and metric preservation do not work in the classical case. Applications of the theorems in statistics, topological phase transition, ergodic theory, etc. are discussed. 
\end{abstract}

\pacs{}


\maketitle

In classical physics, Liouville's theorem (LT) asserts that distribution function $\rho$ in phase space ${\lbrace}p,q{\rbrace}$ is constant along evolution trajectories. \cite{Statistics, Arnold} Intuitively speaking, whether an object is soft or hard in real space, it is equally incompressible in ${\lbrace}p,q{\rbrace}$. If the volume in real space ${\Delta}q$ is changed, its momentum volume ${\Delta}p$ will adjust to maintain ${\Delta}p{\cdot}{\Delta}q$ constant. LT reveals the elegance of classical dynamics (which is largely concealed in Newtonian and Lagrangian mechanics \cite{Arnold}) and serves as a cornerstone of statistical mechanics \cite{Statistics}.

Formally speaking, classical LT is about rigorous incompressibility in phase space ${\lbrace}p,q{\rbrace}$, given Hamilton's equations and local probability conservation; its quantum generalization, Wigner flow \cite{Wigner}, is compressible \cite{Wigner,Flow}, though, except for situations such as free particles, a perfect Harmonic potential. In this work, we discover a pathway to establish a rigorous quantum LT based on Haar measure \cite{Ebook,JoyHaar}, while the previous symplectic measure $dp{\wedge}dq$ \cite{Arnold} proves inappropriate and replaceable.

This work develops arguments in three steps. First, define incompressibility and measure-preservation; show their equivalence (Th. 1, for both classical and quantum),  such that building incompressibility is converted to seeking invariant measure. Second, present measure-preserving theorem (quantum Liouville theorem) and metric-preserving theorem (Th. 2, 3, only for quantum). Third, note theorems’ values in non-equilibrium \cite{Rev,Topoph,ChiragPRL,1, 2, 3,4}, topological transition \cite{TKNN,FuInv,FieldTI,Symm,TPT,ChiragPRX,Song, LiangNM, Bing}, gap-less problems \cite{Son,Sachdev}, strong-interaction \cite{DMFT}, Floquet systems \cite{GenFTh,D15,Baths15,R17}, ergodic theory \cite{Ebook,Song}.

Intuitively, measure $m$ is length, area, volume of a space, depending on the dimension. Formally, measure is a function $m:\mathfrak{B}_{X}{\rightarrow}\mathbb{R}$ based on topological space $X$, which gives the ``volume" of arbitrary (measurable) subsets of $X$ \cite{Ebook,JoyHaar}. $\mathfrak{B}_{X}$ is $\sigma$-algebra of $X$, i.e., a set of subsets of $X$ that satisfies: (i) $X{\in}\mathfrak{B}_{X}$, (ii) if $B{\in}{\mathfrak{B}_{X}}$, $X/B{\in}{\mathfrak{B}_{X}}$, (iii) if $B_n{\in}{\mathfrak{B}_{X}}$,${\bigcup}^{\infty}_{n=1}B_n{\in}{\mathfrak{B}_{X}}$ to rule out non-measurable subsets (e.g., Vitali set \cite{JoyHaar}).

\textbf{Definition 0}. Dynamic evolution is transformation group ${\lbrace}T_t|t{\in}\mathbb{R}{\rbrace}$ formed by inversible (bijective) maps $T_t:X{\rightarrow}X$, where $X$ is a topological space, and a point in $X$ represents a physical state.

\textbf{Remarks}. Def. 0 is for both classical and quantum, as evolutions in both cases are inversible. Hilbert space $\mathcal{H}$ is a vector space, more than a topological space.

\textbf{Definition 1}. Measure preservation (or invariance) is a property of a measure function $m:\mathfrak{B}_{X}{\rightarrow}\mathbb{R}$, namely
\begin{equation}
\begin{split}
m(T(B))=m(B),~{\forall}T{\in}{\lbrace}T{\rbrace},~{\forall}B{\in}\mathfrak{B}_{X},
\end{split}
\label{eq1}
\end{equation} 
associated with dynamic evolution ${\lbrace}T_t|t{\in}\mathbb{R}{\rbrace}$. Measure function $m$ can be expressed in local differential forms
\begin{equation}
\begin{split}
m~{\sim}~{\mu}(x^1,{\ldots},x^N){\cdot}dx^1{\wedge}{\ldots}{\wedge}dx^N,
\end{split}
\label{eq2}
\end{equation}
where ${\mu}(x^1,{\ldots},x^N)$ is a function $X{\rightarrow}{\mathbb{R}}$ and $\wedge$ is the exterior product. Since $\mu$ is another form of $m$, we call both of them measure functions. $x^1,{\ldots},x^N$ are coordinates in $N$-dimension space $X$. For example, in a two-dimension phase space ${\lbrace}p,q{\rbrace}$, symplectic measure has $(x^1,x^2)=(p,q)$ and ${\mu}(p,q)=1$; in the space of SO(3) group ($N=3$), it has $(x^1,x^2,x^3)=({\phi},{\theta},{\psi})$ and Haar measure function
${\mu}({\phi},{\theta},{\psi})=\text{sin}(\theta)$ \cite{JoyHaar}, where ${\phi},{\theta},{\psi}$ are Euler angles. Measure transformation is defined by
\begin{equation}
\begin{split}
{\mu}(x^1,{\ldots},x^N)dx^1{\wedge}{\ldots}{\wedge}dx^N={\mu}^{\prime}(x^{1{\prime}},{\ldots},x^{N{\prime}})dx^{1{\prime}}{\wedge}{\ldots}{\wedge}dx^{N{\prime}}.
\end{split}
\label{eq3}
\end{equation}
$x^1,{\ldots},x^N$ and $x^{1{\prime}},{\ldots},x^{N{\prime}}$ stand for two sets of coordinates. The counterpart of invariance Eq.~\ref{eq1} in $\mu$ is
\begin{equation}
\begin{split}
{\mu}(x^1,{\ldots},x^N)={\mu}^{\prime}(x^1,{\ldots},x^N)
\end{split}
\label{eq4}
\end{equation}

\textbf{Definition 2}. Incompressibility is a property associated with a specific measure function $\mu$ and dynamic evolution ${\lbrace}T_t|t{\in}\mathbb{R}{\rbrace}$ (Def. 0) that for \textit{arbitrary} distribution function ${\rho}(x^1,{\ldots},x^N;t)$,
\begin{equation}
\begin{split}
{\rho}(x^1,{\ldots},x^N;t)={\rho}(x^{1{\prime}},{\ldots},x^{N{\prime}};t^{\prime}),
\end{split}
\label{eq5}
\end{equation}
where $x^i=x^i_{{\lbrace}(x^j)_0{\rbrace}}(t)$ and $x^{i{\prime}}=x^i_{{\lbrace}(x^j)_0{\rbrace}}(t^{\prime})$. $x^i_{{\lbrace}(x^j)_0{\rbrace}}(t)$ are solutions of equations of motion subject to initial conditions ${\lbrace}(x^j)_0{\rbrace}:={\lbrace}(x^1)_0,{\ldots},(x^N)_0{\rbrace}$. Equivalently, we may introduce a single-variable function
\begin{equation}
\begin{split}
{\rho}_{{\lbrace}(x^j)_0{\rbrace}}(t):={\rho}(x^1_{{\lbrace}(x^j)_0{\rbrace}}(t),{\ldots},x^N_{{\lbrace}(x^j)_0{\rbrace}}(t);t).
\end{split}
\label{eq6}
\end{equation}
Incompressibility is expressed as
\begin{equation}
\begin{split}
\dot{\rho}_{{\lbrace}(x^j)_0{\rbrace}}(t)=0.
\end{split}
\label{eq7}
\end{equation}
The time-dependent function ${\rho}_{{\lbrace}(x^j)_0{\rbrace}}(t):{\mathbb{R}}{\rightarrow}{\mathbb{R}}^{+}$ stands for probability density in the vicinity of system setting out from initial states ${\lbrace}(x^j)_0{\rbrace}$. The equivalence  of Eq.~\ref{eq5},\ref{eq7} is evident. We simply plug in Eq.~\ref{eq5} with ${\forall}t,t^{\prime},t^{{\prime}{\prime}}{\ldots}{\in}{\mathbb{R}}$.
The equality holds for arbitrary time, which is exactly $\dot{\rho}_{{\lbrace}(x^j)_0{\rbrace}}(t)=0$. 

Now we present Th. 1: equivalence between incompressibility and measure-preserving.

\textbf{Theorem 1}. Impressibility subject to measure function $\mu$ and dynamic evolution ${\lbrace}T_t|t{\in}\mathbb{R}{\rbrace}$ defined on space $X$ is equivalent to the measure function $\mu$ being invariant under ${\lbrace}T_t|t{\in}\mathbb{R}{\rbrace}$.

\textbf{Proof}. The distribution function evolves with local probability conservation (Appx. A)
\begin{equation}
\begin{split}
{\rho}&(x^1,{\ldots},x^N;t)={\int}{\rho}((x^1)_0,{\ldots},(x^N)_0;0)\\
&{\cdot}{\prod}^N_i{\delta}(x^i-x^i_{{\lbrace}(x^j)_0{\rbrace}}(t)){\cdot}{\mu}_0(x^1_{{\lbrace}(x^j)_0{\rbrace}},{\ldots},x^N_{{\lbrace}(x^j)_0{\rbrace}})\\
&{\cdot}dx^1_{{\lbrace}(x^j)_0{\rbrace}}{\wedge},{\ldots},{\wedge}dx^N_{{\lbrace}(x^j)_0{\rbrace}}.
\end{split}
\label{eq8}
\end{equation}
Eq.~\ref{eq8} is generic evolution, either incompressible or not. It links ${\rho}(x^1,{\ldots},x^N;t)$ with ${\rho}(x^1,{\ldots},x^N;0)$. We have defined time-dependent measure ${\mu}_t(x^1,{\ldots},x^N):={\mu}_0(x^1_{{\lbrace}(x^j)_0{\rbrace}}(-t),{\ldots},x^N_{{\lbrace}(x^j)_0{\rbrace}}(-t)) {\cdot}\mathcal{J}_t(x^1,{\ldots},x^N)$, 
where $ {\mu}_0={\mu}_{t=0}$ and $\mathcal{J}_t:={\partial}((x^1)_0{\ldots}(x^N)_0)/{\partial}(x^1{\ldots}x^N)$ is Jacobian matrix linking two sets of coordinates. Measure preserving Eq.~\ref{eq1},\ref{eq4} give
\begin{equation}
\begin{split}
&{\mu}_0(x^1_{{\lbrace}(x^j)_0{\rbrace}}(t),{\ldots},x^N_{{\lbrace}(x^j)_0{\rbrace}}(t)){\cdot}dx^1_{{\lbrace}(x^j)_0{\rbrace}}(t){\wedge}{\ldots}{\wedge}dx^N_{{\lbrace}(x^j)_0{\rbrace}}(t)\\
&={\mu}_0((x^1)_0,{\ldots},(x^N)_0){\cdot}d(x^1)_0{\wedge}{\ldots}{\wedge}d(x^N)_0
\end{split}
\label{eq9}
\end{equation}
Take derivative of Eq.~\ref{eq8} to estimate ${\partial}_t{\rho}(x^1,{\ldots},x^N;t)$, and  Eq.~\ref{eq9} into ${\partial}_t{\rho}$
\begin{equation}
\begin{split}
&{\partial}_t{\rho}(x^1,{\ldots},x^N;t)=\\
&{\sum}^N_k{\int}{\rho}((x^1)_0,{\ldots},(x^N)_0;0){\cdot}{\partial}_t[{\delta}(x^k-x^k_{{\lbrace}(x^j)_0{\rbrace}}(t))]\\
&{\cdot}{\prod}^{N-1}_{i{\neq}k}{\delta}(x^i-x^i_{{\lbrace}(x^j)_0{\rbrace}}){\cdot}{\mu}_0((x^1)_0,{\ldots},(x^N)_0)\\
&{\cdot}d(x^1)_0{\wedge}{\ldots}{\wedge}d(x^N)_0
\end{split}
\label{eq10}
\end{equation}
Then, apply the chain rule
\begin{equation}
\begin{split}
&{\sum}^N_k{\int}{\rho}((x^1)_0,{\ldots},(x^N)_0;0){\cdot}{\partial}_{x^k}[{\delta}(x^k-x^k_{{\lbrace}(x^j)_0{\rbrace}}(t))]\\
&{\cdot}\frac{d}{dt}(x^k-x^k_{{\lbrace}(x^j)_0{\rbrace}}(t)){\cdot}{\prod}^{N-1}_{i{\neq}k}{\delta}(x^i-x^i_{{\lbrace}(x^j)_0{\rbrace}}(t))\\
&{\cdot}{\mu}_0((x^1)_0,{\ldots},(x^N)_0){\cdot}d(x^1)_0{\wedge}{\ldots}{\wedge}d(x^N)_0=\\
&-{\sum}^N_k\dot{x}^k_{{\lbrace}(x^j)_0{\rbrace}}(t){\cdot}{\partial}_{x^k}{\int}{\rho}((x^1)_0,{\ldots},(x^N)_0;0)\\
&{\cdot}{\prod}^N_i{\delta}(x^i-x^i_{{\lbrace}(x^j)_0{\rbrace}}(t)){\cdot}{\mu}_0((x^1)_0,{\ldots},(x^N)_0){\cdot}d(x^1)_0{\wedge}{\ldots}{\wedge}d(x^N)_0\\
&=-{\sum}^N_k\dot{x}^k_{{\lbrace}(x^j)_0{\rbrace}}(t){\partial}_{x^k}{\rho}(x^1,{\ldots},x^N;t).
\end{split}
\label{eq11}
\end{equation}
Thus, we deduce incompressibility (Def. 2)
\begin{equation}
\begin{split}
\dot{\rho}_{{\lbrace}(x^j)_0{\rbrace}}(t)={\partial}_t{\rho}+{\sum}^N_k\dot{x}^k_{{\lbrace}(x^j)_0{\rbrace}}(t){\partial}_{x^k}{\rho}=0
\end{split}
\label{eq12}
\end{equation}

To show the inverse, we simply need to reverse the derivation from the last to the beginning. Thus, incompressibility and measure-preserving are equivalent. $\square$

\textbf{Remarks}. In Th. 1, $\rho$ is arbitrary, while $\mu$ is particular; that is, incompressibility is a property of a given $\mu$ and ${\lbrace}T_t|t{\in}\mathbb{R}{\rbrace}$, not of $\rho$. Without measure invariance, ${\partial}_t{\rho}$ will contain extra factors (compared with Eq.~\ref{eq11}), and incompressibility is false (Appx. A). Th. 1 converts incompressibility to seeking invariant measure given evolution transformations ${\lbrace}T_t|t{\in}\mathbb{R}{\rbrace}$. Then, we may take advantage of arguments in measure theory \cite{Ebook, JoyHaar} (e.g., existence and uniqueness of invariant measure/metric). This leads to the second step: build quantum LT.

We recap two conditions for classical LT: local probability conservation (continuity condition) and Hamilton equations. Continuity gives
\begin{equation}
\begin{split}
\dot{\rho}_{p,q}(t)={\partial}_t{\rho}(p,q;t)+{\partial}_p{\rho}(p,q;t){\cdot}\dot{p}_{p,q}(t)+{\partial}_q{\rho}(p,q;t){\cdot}\dot{q}_{p,q}(t).
\end{split}
\label{eq13}
\end{equation}
Continuity ${\partial}_t{\rho}(\textbf{r};t)=-{\nabla}{\cdot}(\textbf{J}(\textbf{r};t))=-{\nabla}{\cdot}({\rho}(\textbf{r};t)\textbf{v}(\textbf{r};t))$. Here, spatial coordinates $\textbf{r}{\rightarrow}(p,q)$, and ${\nabla}{\rightarrow}\hat{e}_p{\partial}_p+\hat{e}_q{\partial}_q$. Current density becomes
\begin{equation}
\begin{split}
\textbf{J}(p,q;t)={\rho}(p,q;t)(\dot{p}_{p,q}(t)\hat{e}_p+\dot{q}_{p,q}(t)\hat{e}_q).
\end{split}
\label{eq14}
\end{equation}
Plug Eq.~\ref{eq14} to Eq.~\ref{eq13}
\begin{equation}
\begin{split}
\dot{\rho}_{p,q}(t)=-{\rho}(p,q;t)({\partial}_p\dot{p}_{p,q}(t)+{\partial}_q\dot{q}_{p,q}(t)).
\end{split}
\label{eq15}
\end{equation}

The second condition is satisfaction of Hamilton's equations: $\dot{q}_{p,q}(t)={\partial}_pH(p,q)$ and $\dot{p}_{p,q}(t)=-{\partial}_qH(p,q)$. Combined with Eq.~\ref{eq15}, we have $\dot{\rho}_{p,q}(t)=0$. 

Quantum generalization by Wigner inherits phase space ${\lbrace}p,q{\rbrace}$, \cite{Wigner} although uncertainty principle casts doubt on this notion. Given that Hamilton's equations (HE) are substituted by Schr\"odinger equation (SE), the hope is that by judicious maps (e.g., Wigner function Eq.~\ref{eq16}), incompressiblility should remain. 
\begin{equation}
\begin{split}
{\rho}_W(p,q)=\frac{1}{2\pi}{\int}dq^{\prime}{\varphi}^*(q-\frac{\hbar}{2}q^{\prime})e^{-iq^{\prime}p}{\varphi}(q+\frac{\hbar}{2}q^{\prime}).
\end{split}
\label{eq16}
\end{equation}
The wave function $\varphi(q)$ is mapped to ${\rho}_W(p,q)$ \cite{Note1}. By evaluating the partial derivative ${\partial}_t{\rho}_W(p,q)$ combined with SE, one obtains \cite{Note2}
\begin{equation}
\begin{split}
{\partial}_t{\rho}_W=-{\partial}_q{\rho}_W\dot{q}+{\sum}^{odd}_{\lambda}\frac{(\hbar/2i)^{\lambda-1}}{{\lambda}!}\frac{{\partial}^{\lambda}V}{{\partial}q^{\lambda}}\frac{{\partial}^{\lambda}{\rho}_W}{{\partial}p^{\lambda}},
\end{split}
\label{eq17}
\end{equation}
where $\lambda$ goes over all odd integers. Given terms of $\lambda{\geq}3$ all vanish, Eq.~\ref{eq17} leads to ${\partial}_t{\rho}_W=-{\partial}_q{\rho}_W\dot{q}-{\partial}_p{\rho}_W\dot{p}$, i.e., $\dot{\rho}_W=0$. However, this relies on ${\partial}^{\lambda}V/{\partial}q^{\lambda}=0$ for $\lambda{\geq}3$. In other words, incompressibility only holds for perfect Harmonic oscillators. Thus, LT is only true for classical not for quantum.

Recall that classical LT involves:
\vskip 0.5mm
(1) Dynamics are formulated with a group of inversible maps on a topological phase space $X$.  
\vskip 0.5mm
(2)	Topological space $X$ is phase space ${\lbrace}p_i,q_i{\rbrace}_N$.
\vskip 0.5mm
(3) Incompressibility is linked to symplectic measure ${\prod}_idp_i{\wedge}dq_i$ equipped on $X$.
\vskip 0.5mm
(4)	Robustness: incompressibility is derived \textit{only} from local probability conservation and equations of motion.

Wigner’s generalization inherits (2) and (3), but modifies (4) by replacing Hamilton equations by Schr\"odinger equations (as equations of motion). However, (4) is still violated, as incompressibility further relies on potentials. Point (1) needs more remarks. Quantum mechanics is established on Hilbert space $\mathcal{H}$, more than a topological space like ${\lbrace}p_i,q_i{\rbrace}_N$. The magnitude of a state vector in $\mathcal{H}$ stands for probability, and superposition of two state vectors yields another. However, for 2D ${\lbrace}p,q{\rbrace}$, say $p=0$, $q=0$, $||(p,q)||=0$, which does not mean zero probability; it is meaningless to add two points: $(p,q)+(p^{\prime},q^{\prime})$. Thus, quantum fails point (1). Wigner recovered (1) by introducing Wigner function (Eq.~\ref{eq16}), which transcribes $\mathcal{H}$ to ${\lbrace}p_i,q_i{\rbrace}_N$ and justifies ${\lbrace}p_i,q_i{\rbrace}_N$ despite uncertainty principle. The transformation is inversible (no loss of information), \cite{Wigner} wavefunction is “holographic” in distribution function ${\rho}_W$ on ${\lbrace}p_i,q_i{\rbrace}_N$. It also casts equations of motion into phase space, known as Moyal brackets \cite{Treatise}.

Wigner hoped to achieve a quantum analogue with minimum modifications: maintaining (1)(2)(3) (in fact, partially for (1)), while sacrificing (4). However, the scheme relies on potentials and encounters problems in negative probability, \cite{Wigner,Treatise}, quantization \cite{Quant}, etc. 
We present a different pathway. The proposal is aligned with Wigner’s spirit of mapping $\mathcal{H}$ to topological space \cite{Wigner,Flow,Treatise}, but ${\lbrace}p_i,q_i{\rbrace}_N$ is no longer the choice since uncertainty is averse to it. Additionally, just like classical LT arises from Hamilton's equations, quantum LT should directly arise from Schr\"odinger equation, without referring to Hamiltonian’s forms, to respect point (4) --- robustness. Hence, we inherit points (1)(4), while modify (2)(3): for (2), $\mathcal{H}$ is mapped to the unitary group’s parameter space instead of ${\lbrace}p_i,q_i{\rbrace}_N$; for (3), symplectic measure is replaced with Haar measure. Remember measure function $\mu$ is external equipment (thus to be chosen), rather than intrinsic for a space.

\textbf{Theorem 2}. If the evolution ${\lbrace}T_t|t{\in}{\mathbb{R}}{\rbrace}$ is a unitary group $G$, the distribution function ${\rho}_H$ defined topological space of $G$ equipped with its Haar measure is constant along the evolution trajectory. 

\textbf{Proof}. We need two conditions. The continuity Eq.~\ref{eq14} is now replaced by more general Eq.~\ref{eq8}, and ${\partial}_t{\rho}=-{\nabla}{\cdot}\textbf{J}$ is about choosing symplectic measure ${\mu}{\equiv}1$. The second condition, equations of motion, i.e., SE, enters via evolution $T_t=U(t,0)$ being unitary. That is, ${\lbrace}T_t|t{\in}\mathbb{R}{\rbrace}$ must belong to unitary group, which is compact group. To be specific, two features of SE ensure $T_t$ to be unitary: SE takes the form of diffusion equation (with imaginary coefficients), and Hamiltonian operator is Hermitian: $H=H^{\dagger}$.

Since Haar measure uniquely exists, i.e., ${\mu}_t={\mu}_0$ for ${\lbrace}U(t,0)|t{\in}\mathbb{R}{\rbrace}$, we may always find a unique Haar measure ${\mu}_H$ invariant for $U(t,0)$. Th. 1 states that invariant measure is equivalent to the invariant distribution function ${\rho}_H$. Quantum LT can be expressed as
\begin{equation}
\begin{split}
\dot{\rho}_H(t)=0,
\end{split}
\label{eq18}
\end{equation}
where the subscript ``$H$" refers to the density based on Haar measure ${\mu}_H$, to distinguish from the density (quasi-probability) defined by Wigner function, which has $\dot{\rho}_W(t){\neq}0$. $\square$

\textbf{Remarks}. Classical LT is proved by explicitly finding \textbf{J} (Eq.~\ref{eq14}) based on determined space ${\lbrace}p,q{\rbrace}$ and measure $dp{\wedge}dq$. Quantum LT is based on Haar measure, which refers to a class of measures, i.e., it varies with groups and for each \textit{compact} group it is unique (either analytically or numerically achievable) \cite{JoyHaar}. Proof of quantum LT involves generic argument about Haar measure, rather than referring to specific coordinates or spaces. Similar to classical, quantum LT only relies on two conditions: equations of motion, i.e., SE, and local probability conservation. The former enters via $T_t=U(t,0)$; the latter is via Eq.~\ref{eq8}.


\textbf{Theorem 3}. The metric in group parameter space will remain constant during evolution. 

\textbf{Remarks}. Metric is the “distance” between two points. Th. 3 holds because there always exists (a possibly non-unique) invariant metric associated with an invariant Haar measure. Thus, the measure and metric (volume and distance) will be simultaneously respected. Refer to Chap. 8 of \cite{JoyHaar} for existence proof of invariant metric.

\begin{figure}
\includegraphics[scale=0.9]{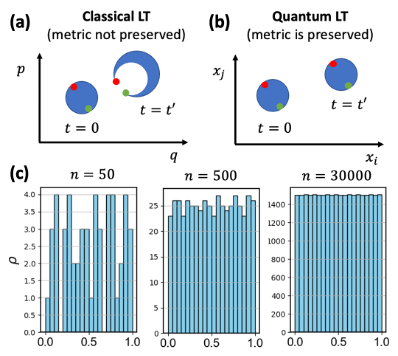}
\caption{\label{fig:epsart}(color online): (a) Classical preserves volume but not distance (e.g., red and green points), like kneading dough. (b) Quantum preserves both volume and distance. (c) At $t{\rightarrow}{\infty}$, $\rho{\rightarrow}$ constant against normalized coordinates (in achievable regions). ${\Phi}={\Omega}=0, {\Theta}={\pi}/4$ (Appx. A) are chosen with a bin number of 20.\label{f1}}
\end{figure}

Why is invariant metric absent in classical dynamics? (Fig.~\ref{f1}(a)(b)) Why cannot Haar measure approach be applied to ${\lbrace}p,q{\rbrace}$? To clarify these questions, we need to see there are some physical principles.
\vskip 0.5mm
(a) Dynamics are formulated with a group of inversible maps ${\lbrace}T{\rbrace}$ on a topological phase space $X$.  
\vskip 0.5mm
(b)	The topological space $X$ on which ${\lbrace}T{\rbrace}$ is defined should be the physical space, i.e., a point in $X$ stands for a physical state.
\vskip 0.5mm
(c) Measure function equipped on $X$ is invariant with ${\lbrace}T{\rbrace}$, i.e., $m(T(B))=m(B)$ for ${\forall}B{\in}\mathfrak{B}_X, {\forall}T{\in}{\lbrace}T{\rbrace}$.

Haar measure is a “special” invariant measure subject to the constraint ${\lbrace}T{\rbrace}=X=G$, which leads to its uniqueness. If we define Haar measure on ${\lbrace}p,q{\rbrace}$, every point $(p,q)$ should stand for a transformation of $G$, leading to $G$ no more than a translation group. That means point $(p,q)$ stands for translating an arbitrary point $(p_0,q_0)$ to $(p_0+p,q_0+q)$; to be invariant Haar measure is ${\mu}_H{\equiv}1$. But do not think ${\mu}_H(p,q){\equiv}1$ is the same as symplectic measure ${\mu}(p,q){\equiv}1$. The difference is that Haar measure has ${\lbrace}T{\rbrace}$ to be translations, totally irrelevant to system’s evolution (principle (a) above is violated). The symplectic measure’s ${\lbrace}T{\rbrace}$ is determined by Hamilton's equations, expressed by the map $p=p_{p_0,q_0}(t), q=q_{p_0,q_0}(t)$. Thus, it is not the case that Haar measure cannot be imposed on classical space ${\lbrace}p,q{\rbrace}$, but rather that principle (a) cannot be simultaneously respected. Quantum LT is non-trivial not only for being rigorously incompressible (principle (c)), but also that principle (a), (b), and (c) can simultaneously be fulfilled. 

\begin{table}
\caption{\label{tab:table1} Measure/metric invariance for Wigner flow, classical/quantum LT, and their valid conditions. HE (SE) means obeying Hamilton's (Schr\"odinger) equations \label{tab1}}
\begin{ruledtabular}
\begin{tabular}{c c c c}
 & Measure & Metric & Valid Cond. \\
\hline
Wigner Flow & no ($dp{\wedge}dq$) & no  & $\partial_q^nV=0$ for $n>2$ \\
Classical LT & yes ($dp{\wedge}dq$) & no & HE \\
Quantum LT & yes (Haar) & yes & SE \\
\end{tabular}
\end{ruledtabular}
\end{table}

Finally, we note applications of the theorems. It transpires that LT’s classical implications \cite{Statistics} can be transplanted to quantum, like uniform distribution $\rho$ over equal-energy surface in ${\lbrace}p_i,q_i{\rbrace}_N$. We plot the counterpart in parameter space (Fig.~\ref{f1}(c)). We use a spin model in a cyclic evolving magnetic field, whose evolution operator $\mathcal{U}$ is given in Appx. A (also Eq. 3 of \cite{Song}). We consider $\mathcal{U}^n$ with $n{\rightarrow}{\infty}$. Note that the cyclic Hamiltonian $H$ is merely for demonstrating; incompressibility is independent of being cyclic.

Incompressibility is required to apply the theorems and tools developed in ergodic theory \cite{Ebook}. Quantum LT offers a valuable non-perturbation approach to exploring areas where perturbation is invalid, such as at topological phase transition (gap closing) \cite{TPT,Song,LiangNM,ChiragPRX, j1}, at quantum critical point \cite{Son,Sachdev}, or electrons with strong interaction or correlation \cite{DMFT} or driven by strong or fast ultra-fast lasers \cite{Rev,Topoph,ChiragPRL,Song23,Yang, 1,2,3,4}. Refer to Appx. B for theorem's application, interpretation, and experimental observation.

The present argument can be extended to infinite dimensions. The trick is expressing infinite $X$ as a product of one finite and one infinite space. Then project dynamics into the finite-dimension quotient space. Extending to infinite space is crucial; position operator is infinite-dimension operator and plays central roles in transport theory \cite{Vanderbilt,Blount}.

\textbf{Conclusion}. We have demonstrated the equivalence between “incompressibility” and “measure-preservation” (Th. 1); prove a quantum Liouville's theorem (Th. 2, Eq.~\ref{eq18}) and metric theorem (Th. 3), confirmed by numerical results (Fig.~\ref{f1}(c)). Liouville's theorem is now rigorously true for both classical and quantum, independent of Hamiltonians (whether $H$ is interacting or time dependent). Liouville's theorem arises from two conditions: local probability conservation and (classical or quantum) equations of motion, while distinctions are highlighted in Table I. Quantum Liouville's theorem provides precise non-perturbation arguments, useful in numerous research fields.

\input acknowledgement.tex   


\section{Appendix A: Proof Notes.}
The 2D phase space ${\lbrace}p,q{\rbrace}$ is used to demonstrate the meaning of generic denotations. Here, $p_{p_0,q_0}(t)$ and $q_{p_0,q_0}(t)$ give the evolution trajectory from initial conditions $p_0,q_0$. For example, for the 1D Harmonic oscillator (whose phase space is 2D, ${\omega}=\sqrt{k/m}$),
\begin{equation}
\begin{split}
&p_{p_0,q_0}(t)=\text{cos}({\omega}t)p_0-m{\omega}\text{sin}({\omega}t)q_0\\
&q_{p_0,q_0}(t)=\text{cos}({\omega}t)q_0-\frac{1}{m{\omega}}\text{sin}({\omega}t)p_0
\end{split}
\label{eq19}
\end{equation}
Conversely, we can express $p_0,q_0$ in $p,q$ and $t$: $p_0=p_{p,q}(-t)$,$q_0=q_{p,q}(-t)$. Then we may evaluate the $2{\times}2$ Jacobian matrix $\mathcal{J}^{(p_0,q_0)}_t(p,q)$ to link the integrals under two distinct sets of coordinates. 

$dx^1_{(x^j)_0}(t){\wedge}{\ldots}{\wedge}dx^N_{(x^j)_0}(t)$ and $d(x^1)_0{\wedge}{\ldots}{\wedge}d(x^N)_0$ in Eq.~\ref{eq9} become $dp_{p_0,q_0}(t){\wedge}dq_{p_0,q_0}(t)$ and $dp_0{\wedge}dq_0$. Note that we do \textit{not} take $dx^i_{(x^j)_0}(t)$ as the time derivative of function $x^i_{(x^j)_0}(t)$ because $t$ is just label of the set of $N$ coordinates. That is, $dx^1_{(x^j)_0}(t){\wedge}{\ldots}{\wedge}dx^N_{(x^j)_0}(t)$ is holistic, and a single $dx^i_{(x^j)_0}(t)$ term is meaningless.

How is Eq.~\ref{eq8} obtained? Consider a single particle moves in space: ${\rho}={\prod}^N_i{\delta}(x^i-x^i_{{\lbrace}(x^j)_0{\rbrace}}(t))$. If it follows a probability distribution (or if we have a swarm of particles), we make a weighted superposition: ${\rho}={\int}{\rho}_0((x^1)_0{\ldots(x^N)_0}){\prod}^N_i{\delta}(x^i-x^i_{{\lbrace}(x^j)_0{\rbrace}}(t))$. Finally, if the space is of non-uniform measure (just imagine gravitational force might distort Euclidean space), we need to further multiply a local measure field $\mu_0$ and finally obtain Eq.~\ref{eq8}. Physically, Eq.~\ref{eq8} is generic expression of local probability conservation (for arbitrary coordinates). If $\mu$ is constant, we obtain familiar Eq.~\ref{eq15}.

Another common mistake is confusion of ${\rho}(x^1,{\ldots},x^N)$ with ${\rho}_{{\lbrace}(x^j)_0{\rbrace}}(t)$: the former is multiple-variable; the latter is single-variable (time). A partial derivative may only act on the former, and the time derivative (e.g., $\dot{\rho}$) only acts on the latter. Recognizing them is crucial for proof derivation. 

Measure invariance (Eq.~\ref{eq1},\ref{eq4}) is indispensable for the proof. If invariance is lacking, we shall obtain the following
\begin{equation}
\begin{split}
&{\rho}(x^1,{\ldots},x^N;t)=\\
&{\int}{\rho}((x^1)_0,{\ldots},(x^N)_0;0){\cdot}{\prod}^N_i{\delta}
(x^i-x^i_{{\lbrace}(x^j)_0{\rbrace}}(t))\\
&{\cdot}{\mu}_0((x^1)_0,{\ldots},(x^N)_0)\frac{{\mu}_0(x^1_{{\lbrace}(x^j)_0{\rbrace}}(t),{\ldots},x^N_{{\lbrace}(x^j)_0{\rbrace}}(t))}{{\mu}_t(x^1_{{\lbrace}(x^j)_0{\rbrace}}(t),{\ldots},x^N_{{\lbrace}(x^j)_0{\rbrace}}(t))}\\
&{\cdot}d(x^1)_0{\wedge}{\ldots}{\wedge}d(x^N)_0.
\end{split}
\label{eq20}
\end{equation}
As such ${\dot{\rho}}_{\left\{\left(x^j\right)_0\right\}}\left(t\right)\neq0$, i.e., it is compressible. 

An example of invariant measure: Haar measure of SO(3). Rotation links two sets  of coordinates: $d\phi^\prime d\theta^\prime d\psi^\prime=\mathcal{J}{\cdot}\ d\phi\ d\theta\ d\psi$, where Jacobian matrix is  $\mathcal{J}={\partial}(\phi^\prime, \theta^\prime,\psi^\prime)/{\partial}(\phi,\theta,\psi)$. Calculation is somewhat tedious, but
\begin{equation}
\begin{split}
d\phi^\prime d\theta^\prime d\psi^\prime=\frac{\sin{\left(\theta\right)}}{\sin{(\theta^{\prime})}}d\phi\ d\theta\ d\psi
\end{split}
\label{eq21}
\end{equation}
That is, $\mu=\mu^\prime=\sin{\left(\theta\right)}$ and $\mu_H\left(\phi,\theta,\psi\right)=\sin{\left(\theta\right)}$. Mind that invariant measure is different from “uniform” measure, which has $\mu$ constant. In general dimensions, definition is given by Eq.~\ref{eq4}. 

LT indicates that in the long time limit, in the achieved phase space, $\rho$ is constant (ergodicity). Evolution operator is \cite{Song}
\begin{equation}
\begin{split}
\mathcal{U}=\begin{pmatrix} \text{cos}({\Theta}/{2})e^{-i{\Phi}} & -\text{sin}({\Theta}/{2})e^{-i({\Omega}-{\Phi})} \\ \text{sin}({\Theta}/{2})e^{i({\Omega}-{\Phi})} & \text{cos}({\Theta}/{2})e^{i{\Phi}} \end{pmatrix}.
\end{split}
\label{eq22}
\end{equation}
Parameters ${\Phi},{\Theta},{\Omega}$ arise from band parameters: gap, driving frequency, etc. \cite{Song}. However, here we only need to take them as parameters of $H$. The parameters ${\Phi},{\Theta},{\Omega}$ are equivalent to Euler angles (the coordinate transformation can be found in Appx. C of \cite{Song}), and the Haar measure is $\sin{(2\Theta)}d{\Phi}{\cdot}d\Theta{\cdot}d\Omega$. Then we evaluate $\mathcal{U}^n$, with an initial $|0{\rangle}=(1,0)^T$, which gives $|{\varphi}_n({\Phi},{\Theta},{\Omega}){\rangle}:=\mathcal{U}^n|0{\rangle}$. Then let $n\rightarrow\infty$, i.e., $t\rightarrow\infty$. And examine the distribution ${\rho}$ over topological space $X={\lbrace}{\Phi},{\Theta},{\Omega}{\rbrace}$ against Haar measure. (Fig.~\ref{f1}(c))

Here we can see difference between Hilbert space $\mathcal{H}$ and the topological space $X$. $|{\varphi}_n({\Phi},{\Theta},{\Omega}){\rangle}$ is vector in $\mathcal{H}$, i.e., 2D vector on $\mathbb{C}$. However, the three-component $({\Phi},{\Theta},{\Omega})$ cannot be added like a vector (although it is extracted from vectors in $\mathcal{H}$), but rather like points in topological space $X$.

\section{Appendix B: The theorem’s physical interpretation and application}
\textbf{The utility of quantum LT}. Why do we need quantum LT? Does it provide information beyond Schr\"odinger equation (SE)? Although SE and initial conditions carry all dynamic information, it is usually unsolvable. Quantum TL is not to provide “new” information but to access information from SE (e.g., asymptotic, or statistical behavior) without the need to solve it. 

The mechanism is, because ${\rho}_H$ is constant along the achievable regions, if one’s interest is statistical or asymptotic behavior (most observables belong to this type), one may switch from solving the true evolution path in Hilbert space to solving the achievable region in unitary group parameter space against Haar measure. We do not care the temporal order of the system traversing these regions, but only the region, which is a more tractable problem.

In fact, similar strategies have been used by classical LT and classical statistics: it does not matter how the system covers phase space ${\lbrace}p,q{\rbrace}$, but only the achievable regions and the corresponding probability density. As such, solving $N$-particle Hamiltonian equations is eluded, and statistical behaviors of unsolvable large system could be formulated.

\textbf{Application examples} Quantum LT needs some models to demonstrate its power to yield concrete results. Just like when classical LT (occupancy $e^{-{\beta}T}$ deduced from it) is applied to transports, one gets conductivity rules (e.g., temperature dependence); when it is applied to free particle models, one gets dilute gas behaviors.

In spin/band models \cite{Song}, solving the reachable region corresponds to finding ``ergodic subgroup" (Appx. C of \cite{Song}), via which one can obtain analytic solutions of spin or inter-band pumping (quantum LT was then termed as a ``measure-preserving formalism" and is now formalized into Th. 2). Without quantum LT, pumping probability $p_G$ is expressed with an infinite series, evaluated by $\mathcal{U}^n$ with $n{\rightarrow}{\infty}$. $p_n$ exhibits as a complicated series.
\begin{equation}
\begin{split}
&p_1=\text{sin}^2(\frac{\Theta}{2})\\
&p_2=\frac{1}{2}(p_1+|-\text{sin}(\frac{\Theta}{2})\text{cos}(\frac{\Theta}{2})-e^{2i{\Phi}}\text{sin}(\frac{\Theta}{2})\text{cos}(\frac{\Theta}{2})|^2)\\
&p_3=\frac{1}{3}(2p_2+|\text{cos}(\frac{\Theta}{2})(-\text{sin}(\frac{\Theta}{2})\text{cos}(\frac{\Theta}{2})-e^{2i{\Phi}}\text{sin}(\frac{\Theta}{2})\text{cos}(\frac{\Theta}{2}))\\
&-\text{sin}(\frac{\Theta}{2})(-\text{sin}^2(\frac{\Theta}{2})+e^{-2i{\Phi}}\text{cos}^2(\frac{\Theta}{2}))|^2)\\
&p_4=\frac{1}{4}(3p_3+...)
\end{split}
\label{eq23}
\end{equation}
Since $p_G=p_{\infty}$, one can imagine what horrible expression it will be. But with quantum LT, one can prove $p_G$ converge to a compact analytic solution: 
\begin{equation}
\begin{split}
p_G=\frac{\text{sin}^2(\frac{\Phi}{2})}{2(1-\text{cos}^2(\frac{\Phi}{2})\text{cos}^2(\Phi))}.
\end{split}
\label{eq24}
\end{equation}

Via the analytic solution above, one may deduce a concept “geometric pumping” in both spin and band scenarios, \cite{Song} whose defining feature is pumping probability only depends on geometric/topological parameters irrelevant to energetic ones. Thus, quantum LT also helps establish novel physical concepts. 

\textbf{Experiment}. Empirical information can be obtained by testing the phenomena predicted by quantum LT. Quantum LT is not a specific observable but is a law that influences broad phenomena. Take geometric pumping as an example. (It is not the only case, just because a TPT model is solved with the theorem, we use this to demonstrate.)

Passing from classical statistics to Fermion/Boson statistics, a crucial thing is to lower temperature to make quantum effect emerge; we only need to examine ``conventional" observables but in search of abnormality against classical interpretation. Similarly, detecting quantum LT does not need very fancy measurement. A simple path is to perform measurement around topological phase transition (TPT), i.e., at band/level degeneracy. 

For example, we apply quantum LT to a two-band model that undergoes periodic gap closing, we obtain exact analytic solutions. For details of the model and the solving process, one can refer to \cite{Song}. Here we just quote the result: if gap closing changes the topological state of bands, 1/2 electron will be pumped to the upper band at gap closing $k_0$; if gap closing does not alter the topological state, no electron will be pumped. 

To test this prediction by quantum LT, we need (i) a topological insulator at the vicinity of TPT (in practice, that means the insulator’s gap cannot be too large); (ii) a means of realizing periodic TPT; (iii) a means of measuring the amount of pumping charge (hopefully with capability of time resolution). 

For (i), we can choose ZrTe$_5$, a topological insulator ($Z_2$ type) features a single band cone at $\Gamma$ in B.Z. \cite{ChiragPRX, LiangNM}, with gap 10-100 meV. For (ii), we can use phonon to drive the band and periodically close the band gap to realize periodic TPT. In ZrTe$_5$, this can be done by, e.g., $A_{1g}$ phonon mode ($\sim$1.2 THz). For (iii), we can use ultra-fast spectrum or pump-probe techniques \cite{ChiragPRX, LiangNM}). The amount of charge being pumped will be proportional to the change of reflectivity ${\Delta}R$ or transition ${\Delta}E$  (compared with the ground state), i.e., ${\Delta}R$ (or ${\Delta}E$) ${\propto}Q_{pump}$. \cite{ChiragPRX}

Since geometric pumping is fractional and relies on the presence of TPT, we anticipate (a) the pumping might happen even at sub-gap excitation, i.e., driving frequecny is less than average gap $\hbar{\omega}{\ll}\bar{\Delta}$; (b) such pumping disappears given TPT is absent; (c) the saturated $Q_{pump}$ is lower than energetic pumping by quasi-particles.

\textbf{Advantages and limitations of quantum LT compared with classical LT and Wigner function}. Quantum LT arises from modifying the Wigner function, fixing a major shortcoming for Wigner function: loss of ``incompressibility". Moreover, the metric preservation (Th. 3) renders a rigid-body motion for wave package, i.e., for arbitrary wave package, it will not disperse, which provides another pathway of understanding the long lifetime of wave package other than soliton approaches. On the other hand, since Wigner function involves position $q$, it can handle real-space problems, e.g., transport, under a semi-quantum picture ($p,q$ are both definite). The quantum LT currently works for space of finite dimensions, because the dimension $N$ of U($N$) group must be finite. Thus, position-related problem (position is an infinite-dimension quantum operator) is still unachievable for the current version of quantum LT. 

That is why we try to generalize the theorem to infinite dimensions. In the end of main texts, we forecast our next work by saying: “The present argument can be extended to infinite dimensions. The trick is expressing infinite $X$ as a product of one finite and one infinite space…”, which gives a prospective of simultaneously realizing three aspects: (i) quantum, (ii) infinite dimensions of Hilbert space, (iii) incompressibility; in comparison with classical LT only fulfilling (ii)(iii), and Wigner function only fulfilling (i)(ii). It is intriguing to see in what an extent quantum LT can cover Wigner function’s jobs. Nonetheless, the current theorem already shows its advantage in treating quantum degrees of freedom, e.g., spin, inter-band scenarios, which are unachievable for Wigner function. 

\textbf{Does Haar measure have a physical meaning}? In this section, we link Haar measure with the physical probability. In view of the fundamentality and controversy related to quantum interpretation, for example, even the notion “probability” might have different interpretations (ch. 11 of \cite{Qfound}), which lead to different logic lines, we must stress the present concern about Haar measure is merely that it is physically suitable or convenient, and the probability density associated to it means a common sense; a deeper understanding must be established upon the full understanding of quantum measurements \cite{Qfound}, which is apparently beyond the scope of this work.

A physical meaning as “probability density” is usually assigned to a density field; one does not usually give a separate physical meaning to a measure function, as it is an ingredient in defining a density field. It is not always possible to extend the wondering of “what is the physical meaning” along an array of math decomposition. However, as a heuristic, one may imagine measure as a “ruler” to determine a density; different rulers will give different density readings, but the different readings correspond to the same physical state. That is, different measure functions are linked by transformations; choosing different measure function does not alter physics. 

Thus, in a sense, different definitions of density are on the equal status, if one can accept the density defined in real space ${\Delta}q$, or density defined in phase space by ${\Delta}q{\Delta}p$, one should accept the density defined with Haar measure without doubts. Each density corresponds to a particular choice of space and measure function equipped on the space (such a space is called \textit{measure space} in measure theory). 

Given every measure (and associated density) is equal, what makes Haar measure (symplectic measure) uniquely outstanding in quantum (classical)? That is, a good measure function should be a time-independent (static) one that can maintain incompressibility for arbitrary systems, which mean arbitrary initial states, arbitrary potentials, arbitrary number of particles. A poorly selected measure function can possibly maintain incompressibility, but it must adjust its form based on the motion of particles (thus a time-dependent function), which will be trivial. That is why in Eq.~\ref{eq8} the measure function field is a static ${\mu}_0$ rather than ${\mu}_t$. Thus, constructing an incompressible formulation is an incomplete (thus kind of misleading) statement for LT; one should keep in mind of using a static measure function. Understanding these constraints in constructing incompressibility, one is easy to understand why a measure needs to be of very nice mathematical properties, which are not optional or preferred, but mandatory. 

Choosing a good measure function like Haar measure in quantum or symplectic measure in classical helps reveal ``stable points" of a dynamic system (thus leading to stable observables), while a poorly chosen measure will probably give an illusion that the system is still unstable. For example, if we choose a non-uniform measure in ${\lbrace}p,q{\rbrace}$ space, at equilibrium, the density still keeps changing. 

There are infinite ways in defining a density (i.e., choosing a measure function), but there is only a single one among them that correctly reflects physical stable states; and that density (also the associated measure function) can be considered physically meaningful or convenient.

Note that building a complete logic for quantum \cite{Qfound} is not a ground to be covered by this paper. We focus on more certain arguments and achievable aspects. (1) The physical probability is most commonly described by a density field, which is defined upon a specific measure function (Haar measure is one possible choice). (2) A physically suitable and convenient (also mathematically non-trivial) measure function should be a static one and invariant under physical evolution. (3) Based on this criteria, symplectic measure and Haar measure become (potentially uniquely) outstanding for classical and quantum, respectively. (4) Haar measure will help simplify calculations, analytically solve models, allude to novel physical concepts or phenomena, as demonstrated here. These are indicators (although not proof) for Haar measure having physical meanings and significance. 

Now we can connect the above physical discussions back to the formal (maybe a little abstract) presentation in the main texts. As we mentioned, being static boils down to the fact in Eq.~\ref{eq8} the measure field is a static field not being allowed to change over time. Whether it is time-independent, and whether the arbitrariness (in initial conditions, potentials, etc.) is true boil down to transformation properties of measure function (Def. 1 in the paper) under the evolution group (Def. 0 in the paper) determined by the equation of motions in classical or quantum scenarios. 

Fundamentally, these considerations boil down to the three principles (a)(b)(c) raised after the Th. 3. We apply the three principles to an analysis on inapplicability of Haar measure to classical. To be concrete, as we mentioned, the simultaneous satisfaction for the three principles makes $dp{\wedge}dq$ outstanding for classical and Haar measure outstanding for quantum. The reverse inapplicability (symplectic applied to quantum) is exactly the failing of incompressibility for Wigner function.

Back to the question ``whether Haar measure has physical meanings", we make two non-decisive but constructive comments. (1) A physically suitable measure should be invariant under evolution transformation (Def. 0), just like people believe a valid relativistic quantum theory should be invariant under Lorentz group, although reconciling quantum with relativity theory still eludes us. (2) A physical measure should be useful. We have given preliminary evidence for Haar measure, and there are more to explore. For example, if Haar measure is useful in understanding the long lifetime of particles, as the wave package picture suffers from dispersion. Given such evidence is accumulated, one may promote quantum LT (the measure conservation) to a more fundamental state.

\textbf{A guideline for the physics/math background}. To bridge the gap between physics and math, we provide a guideline. There are three tiers. First, for convinced and proficient readers, they can quote the result as 
``The time evolution is unitary and the Haar measure remains invariant under time evolution; the invariance Haar measure implies the corresponding distribution remains constant over time." Second, for readers interested in techniques of the proof, the kernel knowledge includes: (i) the derivation of classical LT, refer to Ch. 1 of Gibbs's book \cite{Gibbs}. (ii) Ch. 1-2 of Bogoliubov’s book \cite{bogoliubov}, because our derivation of Th. 1 could be considered a generalization of Bogoliubov’s proof of classical LT by including a local measure field. (iii) Invariant measure and Haar measure. Most illuminating examples to quickly capture these topics: Haar measure of SU(2) or SO(3) group, the measure function associated to rational and irrational numbers on real number axis, Vitali set as examples to understand non-measurable sets. (iv) Derivation regarding Wigner functions (e.g., Eq.~\ref{eq16},\ref{eq17}) is found in \cite{Wigner}. Third, we also provide references for more ambitious readers, who are aimed to explore uniqueness of Haar measure, \cite{JoyHaar} applications of quantum LT intertwining with ergodic theory \cite{Ebook} or general topics in measure theory.

\end{document}

%% file: acknowledgement.tex
\textbf{Acknowledgement}. This work was supported by the Ames National Laboratory, the US Department of Energy, Office of Science,
Basic Energy Sciences, Materials Science and Engineering Division under contract No. DEAC02-07CH11358. 